%% file: 2024_IEEE_T-IV.tex
\documentclass[journal]{IEEEtran}
\pdfoutput=1
\usepackage{amsmath,amsfonts}
\usepackage{algorithmic}
\usepackage{algorithm}
\usepackage{array}
\usepackage[caption=false,font=normalsize,labelfont=sf,textfont=sf]{subfig}
\usepackage{textcomp}
\usepackage{stfloats}
\usepackage{url}
\usepackage{verbatim}
\usepackage{graphicx}
\usepackage{cite}

\usepackage{doi}
\usepackage{comment}
\usepackage{booktabs}
\usepackage{makecell}
\usepackage{amssymb}
\usepackage[export]{adjustbox}
\usepackage{placeins}
\usepackage{afterpage}
\usepackage{siunitx}
\usepackage{pifont}
\usepackage{amssymb}
\newcommand{\cmark}{\ding{51}}
\newcommand{\xmark}{\ding{55}}
\ifCLASSINFOpdf
\DeclareGraphicsExtensions{.pdf}
\fi
\usepackage{subfig}
\usepackage{multirow}
\usepackage{cite}
\usepackage{orcidlink}
\DeclareMathOperator{\sign}{sign}

\newcommand{\anova}[4]{\mbox{$(F(#1,#2)\!=\!#3,~p\!=\!#4)$}}
\newcommand{\anovaPL}[4]{\mbox{$(F(#1,#2)\!=\!#3,~p\!<\!.001)$}}

\newcommand{\ChiSquare}[3]{\mbox{$(\chi^2(#1)\!=\!#2,~p\!=\!#3)$}}
\newcommand{\ChiSquarePL}[3]{\mbox{$(\chi^2(#1)\!=\!#2,~p\!<\!.001)$}}

\newcommand{\studentTPL}[3]{\mbox{$t(#1)\!=\!#2,~p\!<\!.001$}}
\newcommand{\yuenT}[3]{\mbox{$\mathrm{Yuen's-}t(#1)\!=\!#2,~p\!=\!#3$}}

\newcommand{\mannWhitney}[2]{\mbox{$U\!=\!#1,~p\!=\!#2$}}
\newcommand{\mannWhitneyPL}[2]{\mbox{$U\!=\!#1,~p\!<\!.001$}}

\begin{document}
\title{Steering Feedback in Dynamic Driving Simulators: The Influence of Steering Wheel Vibration \\and Vehicle Motion Frequency}

\author{Maximilian Böhle \orcidlink{0000-0002-2919-000X}, Bernhard Schick \orcidlink{0000-0001-5567-3913} and Steffen Müller
\thanks{Manuscript submitted March 22, 2024}
\thanks{Maximilian Böhle is with the Institute for Driver Assistance and Connected Mobility at Kempten University of Applied Sciences, Kempten, Germany and with the Department of Automotive Engineering at Technical University Berlin, Berlin, Germany(e-mail: maximilian.boehle@hs-kempten.de) \\
Bernhard Schick is with the Institute for Driver Assistance and Connected Mobility at Kempten University of Applied Sciences, Kempten, Germany (e-mail: bernhard.schick@hs-kempten.de) \\
Steffen Müller is with the Department of Automotive Engineering at Technical University Berlin, Berlin, Germany (e-mail: steffen.mueller@tu-berlin.de)}
}

\markboth{PREPRINT, THIS SUBMISSION WAS ACCEPTED FOR PUBLICATION BY THE IEEE TRANSACTIONS ON INTELLIGENT VEHICLES}%
{Boehle \MakeLowercase{\textit{et al.}}: Influence of steering wheel and vehicle \\motion frequency on steering feedback \\in dynamic driving simulators}

\IEEEpubid{This work was accepted for publication by the IEEE. The published version is available at \doi{https://doi.org/10.1109/TIV.2024.3401868}}

\maketitle


\begin{abstract}
The validity of the subjective evaluation of steering feedback in driving simulators is crucial for modern vehicle development. Although there are established objective steering characteristics for the assessment of both stationary and dynamic feedback behaviour, factors such as steering wheel vibrations and vehicle body motion, particularly in high-frequency ranges, present challenges in simulator fidelity. This work investigates the influence of steering wheel vibration and vehicle body motion frequency content on the subjective evaluation of steering feedback during closed-loop driving in a dynamic driving simulator. A controlled subject study with 30 participants consisting of a back-to-back comparison of a reference vehicle with an electrical power steering system on a country road and three variants of its virtual representation on a dynamic driving simulator was performed. Subjective evaluation focused on the representation of road feedback in comparison to the reference vehicle. The statistical analysis of subjective results show that there is a significant influence of the frequency content of both steering wheel torque and vehicle motion on the subjective evaluation of steering feedback in a dynamic driving simulator. The results suggest an influence of frequency content on the subjective evaluation quality of steering feedback characteristics that are not associated with the dynamic feedback behaviour in the context of established performance indicators. 

\end{abstract}


\section{Introduction}
\label{chap_Introduction}
\IEEEPARstart{T}{he} development of modern vehicles relies heavily on driving simulators for human-centered design. Steering feel, especially the feedback torque, plays a pivotal role in subjective vehicle evaluation \cite{rothhamel2011method,harrer2017steering,gomez2015findings,bootz2016fahrwerk,uselmann2017beitrag}. The growing importance of virtual methods in steering system development is putting additional emphasis on this field of research. Despite recent advancements resulting in extensive development work being carried out virtually, substantial parts of steering feel development are still being performed iteratively based on subjective evaluation through system experts in real vehicles \cite{nippold2016analysis,gruener2017objectification,dusterloh2018absicherung,ketzmerick2022validated,haas2023improvements}. Since the trend towards higher levels of automated driving and the increasing market relevance of Steer-by-Wire systems introduces new challenges regarding the safety of classic development processes, there is growing interest in subjective evaluation methods that can transfer steering characteristics from simulators to real vehicles early in the development process \cite{bertollini1999applying,zong2007study,honisch2015verbesserung}.

\subsection{State of research}
\IEEEpubidadjcol
In recent years there has been a large amount of published work investigating the role \cite{liu1995force,mourant2002evaluation,toffin2003influence,toffin2007role,shyrokau2016influence} and validity \cite{salaani2002modeling,iyasere2007real,katzourakis2010steering,baumann2014evaluation,gomez2016validation} of steering feel in driving simulators. Steering feel is comprised of a variety of sensory impressions, of which haptic feedback in the form of steering wheel torque (SWT) and visual and kinesthetic feedback on the response of the vehicle body are generally considered to be decisive \cite{wolf2009ergonomische,barthenheier2007potenzial,harrer2017steering}. Steering feedback from SWT is particularly important for subjective assessment of steering feel and vehicle dynamics \cite{gordon1966experimental,fiala1967,odenthal2003uebertragung,gomez2015findings,uselmann2017beitrag} and has been the focus of the majority of recent works on steering feel in driving simulators. Several simulator studies have concluded that SWT can have a significant effect on drivers' performance of the lateral control task \cite{liu1995force,toffin2003influence,toffin2007role,samiee2015effect,kim2011model} and can influence drivers' subjective assessment of overall vehicle dynamics \cite{mourant2002evaluation,decker2009beurteilung,mandhata2012evaluation}. There are however strong indications in literature that secondary factors such as steering wheel and vehicle body vibrations and acoustic contributions can have a relevant influence on subjective ratings, particularly concerning the perception of road feedback \cite{bellmann2002perception,harrer2007characterisation,zschocke2009beitrag,berber2009evaluation}.

Another focus of research has been the transfer of established development processes into virtual environments. Although several recent works demonstrate the successful industrial implementation of virtual steering system tuning 
\cite{duesterloh2019objectification,grau2018steering,wang2016epas}, there is still evidence that not all relevant influencing factors for the subjective evaluation of steering feedback are known and can be reproduced with sufficient accuracy in driving simulators \cite{gomez2015findings,harrer2017steering,dusterloh2018absicherung,ketzmerick2022validated, haas2023improvements}. Due to the high effort required for the iterative subjective tuning of modern steering systems, the objectification of steering feel has been investigated for decades. A wide range of objective steering feel characteristics have been developed through the correlation of subjective evaluations with objective data obtained from standardized open-loop vehicle dynamics maneuvers through regression models \cite{brunn2004objektivierung,harrer2006steering,harrer2007characterisation,zschocke2008links,wolf2009ergonomische,zschocke2009beitrag,rothhamel2011method,ketzmerick2022validated}. The objective characteristics derived in this way describe the feedback behavior largely in steady-state conditions, for example using the dead zones and amplitude ratios of steering wheel angle and steering wheel torque measured during a weave test. Characterization of the dynamic feedback behavior linked to road feedback typically involves the analysis of the steering system transfer function in the frequency range of up to \SI{30}{\hertz} to identify the system bandwidth via the corner frequency of the amplitude response \cite{brunn2004objektivierung,groll2006modifizierung,pfeffer2011akustik,muenster2014requirements,lunkeit2014beitrag,fankem2014new,uselmann2017beitrag,duesterloh2019objectification}. This frequency range covers all first-order wheel-induced excitations in relevant velocity ranges. However, the range of relevant excitations for the subjective evaluation of steering wheel vibration has been the subject of debate \cite{grau2016objective}. Some works propose that excitations in frequency bands of up to \SI{120}{\hertz} \cite{berber2009evaluation} can be relevant for the identification of road feedback and subjective evaluation \cite{bellmann2001methoden,bellmann2002perception,giacomin2004beyond,berber2009evaluation,ajovalasit2013human,henneberger2016akustische}. 
To summarize, although there are established objective steering characteristics for the assessment of both stationary and dynamic feedback behavior, there are indications that there can be contributors to the subjective evaluation of steering feedback characteristics, such as road feedback, that exceed the known evaluation criteria. 

\subsection{Contribution of present research}
This work contributes to the understanding of the influence of frequency content of steering wheel vibrations and vehicle body motion on the subjective evaluation of steering feedback in dynamic driving simulators. The objective of this study is to identify factors that contribute to differences between the subjective evaluation of steering feedback in the driving simulator and that of the real vehicle, beyond the scope of established objective characteristics.

The main contributions of this work can be summarized as follows:
\begin{itemize}
    \item Conduction of a controlled back-to-back subject study consisting of a drive in a reference vehicle followed by a drive in a validated representation of the reference vehicle in a high-fidelity driving simulator.
    \item Investigation of the effect of SWT frequency content on the subjective evaluation of steering feedback utilizing a high-fidelity representation of the reference system's feedback characteristics.
    \item Investigation of the effect of vehicle body motion frequency content on the subjective evaluation of steering feedback in a dynamic driving simulator.
\end{itemize}

\section{Related work}
Existing work differs substantially from this study, both in terms of methodology and research focus. Although previous work has emphasized the importance of road feedback \cite{mandhata2004investigation,zschocke2008links,zschocke2009beitrag,grau2018steering}, steering wheel and vehicle body vibrations have primarily been investigated as comfort issues, rather than as factors that influence steering feel \cite{plunt1999strategy,bellmann2001methoden,giacomin2003frequency,kaster2005acoustics,amman2005equal,ajovalasit2009non,knauer2010objektivierung,wang2014robust,parduzi2021bewertung}. The underlying assumption that, if relevant, any kind of high-frequency excitation in the steering wheel is a disturbing influence, is not consistent with the findings from steering feel studies. Specifically research on driver perception of road surfaces shows contradicting results \cite{giacomin2004beyond,giacomin2005study,berber2009evaluation}. We are therefore conducting a steering feel evaluation focussing on these feedback characteristics.
Methodologically, this study is distinguished by its back-to-back design enabling subjective evaluation of steering feedback during closed-loop driving in consistent driving conditions in a high-fidelity dynamic driving simulator. A key factor for the feasibility of the presented study is the fidelity of the steering representation in the driving simulator. Several recent works have presented approaches for generating realistic steering feedback on driving simulators, ranging from physical \cite{camuffo2002simulation,salaani2002modeling,mills2003behavioural,inaba2016simulators,certosini2019development} and data-driven models \cite{fankem2014new,van2016practicability,zhao2022modeling,zhao2023improved} to Hardware-in-the-Loop (HiL) setups that allow the feedback of model-based rack forces through the physical steering system \cite{segawa2006preliminary,baumann2014evaluation,honisch2015verbesserung,wang2016epas,nippold2016analysis,vinattieri2016target,vinattieri2017steering,talarico2021virtual}. The majority of model-based approaches highly simplify the software contribution to SWT in electrical power steering (EPS) systems, despite its substantial influence on feedback behavior \cite{groll2006modifizierung,schimpf2016charakterisierung,uselmann2017beitrag,duesterloh2019objectification}. This presents a major conceptual limitation for the investigation of dynamic steering feedback phenomena. Although HiL test benches \cite{kim2002control,lee2011development,lee2012research,schimpf2016charakterisierung,uselmann2017beitrag} have been an integral part of steering system development via objective characteristics, their popularity as driving simulators for subjective evaluation is still limited. For this reason, available studies with a similar research focus do not represent steering feedback of road excitations with sufficient fidelity. The following provides a brief overview of closely related work.

\cite{shyrokau2015effect} investigated the effect of steering model complexity on steering feel evaluation in a static simulator, \cite{shyrokau2016influence} and \cite{shyrokau2018effect} further investigated the influence of driving simulator motion on the subjective evaluation of on-center handling. All three studies utilized a steering system representation with limited fidelity regarding feedback behavior, additionally the characteristics selected for subjective evaluation do not cover the research questions investigated in the present study. \cite{bertollini1999applying} uses a dynamic simulator to investigate the effects of vehicle motion on subjective steering feel evaluation but only compares the variants static and dynamic without further differentiation. Additionally, there is no evaluation of road feedback and no representation of road surface in the simulation environment. \cite{gomez2016validation} investigates the validity of steering feel evaluation in a dynamic simulator but the steering model contains no representation of EPS software outputs and subjective evaluation does not cover road feedback. \cite{rothhamel2011method} investigates subjective steering feel evaluation in a dynamic driving simulator but the simulator only represents motion in three Degrees of Freedom (DOFs) and there is no representation of road surface in the simulation environment. 

\section{Subject study design}
The study consisted of a back-to-back comparison between the reference vehicle and three variants of its representation in a driving simulator. All drivers were subjected to the same general test procedure, the order of variants in the driving simulator was however rotated to mitigate the effect of discussions between participants and confirmation bias through study design. After providing written consent to the participation in the subject study and the subsequent analysis and scientific publication of their evaluation data, the participants received a briefing on the study design and the evaluation criteria before participating in the driving sessions. Each driving session consisted of one lap in the reference vehicle on a public road followed by three laps on the driving simulator. Subjective ratings were collected after each finished lap, both in the reference vehicle and in the driving simulator, resulting in four full sets of evaluation criteria per participant. Each participant's reference lap was preceded with a six kilometer drive from the Institute for Driver Assistance and Connected Mobility (IFM) to the starting point of the reference track. The trip to the starting point took around \SI{8}{\min} one-way and was used as a familiarization drive with the reference vehicle. After completing the evaluation on the reference lap with a length of \SI{11.5}{\kilo\metre}, the measured data were automatically prepared for the simulator drive upon return to the IFM. Participants were then asked to start with their first simulator variant, provided they did not require a break. All participants conducted the study in similar weather conditions on dry roads.

\input{ClassicKPIs.tex}

\subsection{Participants}
The necessary sample size for the described study design was determined with the program G*Power \cite{faul2007gpower}, version 3.1.9.7. An a priori power analysis for the within-factors repeated measures design with the parameters shown in \autoref{tab_GPower} in the Appendix yielded a critical value of \anova{3}{69}{2.737}{0.05} at a sample size of $24$. 

In total, $30$ drivers participated in the study. Four of the participants were female, $26$ male, the average age was $M = 34.7$ with a standard deviation of $SD = 9.29$ years, covering a range from $24$ to $60$. To ensure high evaluation quality, minimize preparation time, and prevent dropouts due to motion sickness, only drivers with extensive experience on the driving simulator were selected. Based on their current and past professional experience, the participants were divided into different user levels ranging from normal driver to steering system expert. There was one dropout due to motion sickness and two due to technical issues with data recording during the study conduction reducing the number of valid datasets to 27. \autoref{tab_UserLevels} in the Appendix shows the definition of user levels and their distribution among datasets. Participation in the study was voluntary and no monetary compensation was provided.

\input{fig_aVDS.tex}

Subjective evaluation of steering feel is typically performed by professional test drivers with comprehensive experience in the specific evaluation of steering feel characteristics and technical expertise regarding the related steering system properties. There are several works showing the limitations of subjective evaluation of steering feel through normal drivers and non-expert evaluators due to the decreased repeatability and resolution capability of overall ratings, narrower vehicle dynamics limits during evaluation or a lack of understanding of specific evaluation criteria \cite{riedel1997subjektive,zschocke2009beitrag,harrer2017steering,shyrokau2018effect}. Other works have however shown that with an adjusted questionnaire in combination with a comprehensive technical briefing, limited evaluation volume through both questionnaire content and overall study duration, subjective evaluation of steering feel through normal drivers is possible \cite{riedel1997subjektive,krueger1999bewertung,zschocke2009beitrag}. In line with these findings, study time was limited to one hour and a specifically designed evaluation catalogue was used.

\subsection{Reference vehicle}
The reference vehicle used was the sports variant of a Golf class vehicle with an EPS system featuring a progressive gear ratio. To maximize the available steering feedback within the limitations of the stock EPS software, the vehicle was equipped with high performance tires and the EPS software was set to dynamic mode. The measurement setup for the reference lap consisted of a combined inertial and gyroscopic measurement platform, a GNSS measurement unit with correction data, strain gauge-based force sensors on both front axle tie rods, and a real-time computer for time-synchronous recording of the vehicle bus signals to and from the EPS ECU. All measurement data were synchronized to use the same time base and resampled using linear interpolation to the global sample rate of \SI{1}{\kilo\hertz}. Since there was no dedicated measurement of vehicle velocity, GNSS data were corrected with IMU measurements before being used in the driving simulator. Prior to the study conduct, a wheel alignment was performed and except for variations in driver weight, the reference vehicle was kept in identical load conditions as during validation drives.

\subsection{Driving simulator}

The simulator drives in this study were performed on the dynamic driving simulator at the IFM of the Kempten University of Applied Sciences. The Advanced Vehicle Driving Simulator (aVDS, see \autoref{fig_aVDS}) features a motion platform with eight electric linear actuators, enabling the representation of vehicle motions with translational accelerations over \SI{10}{\meter/\square\second} and rotational accelerations over \SI{1100}{\degree/\square\second} in all DOFs. Frequency bandwidth is illustrated in \autoref{tab_AVDS_performance}. The simulator incorporates an EPS HiL setup, allowing the simulation of a complete steering system with external rack force feedback. The reference vehicle is represented by a cockpit with a fully functional interior. The simulation environment is visualized using seven laser projectors with a refresh rate of \SI{240}{\hertz} on a \SI{270}{\degree} cylindrical screen measuring \SI{8}{\meter} in diameter and \SI{4}{\meter} in height. The simulation output is received by seven rendering PCs from a synchronization node at \SI{1}{\kilo\hertz}. The vehicle, road, and tire models are running on a real-time PC using RedHawk Linux. Real-time scheduling and monitoring of parallel model execution and IO communication via CAN, UDP, and EtherCAT is performed by a SIMulation Workbench instance running at a sample rate of \SI{1}{\kilo\hertz}. Platform motion is controlled by a separate real-time PC that runs the motion cueing algorithm (MCA) and subsequent motion control at \SI{2}{\kilo\hertz}. The transmission of steering-related data is performed through a synchronized EtherCAT network at \SI{8}{\kilo\hertz}. The resulting physical validity in terms of classic steering feel KPIs is displayed in \autoref{tab_KPIs}.

\input{AVDS_performance}

\subsubsection*{Motion cueing}
In the presented study, the aVDS runs a classic washout MCA without tilt coordination, as described in \cite{nahon1985flight} for all translational DOFs and yaw rotation. Motion inputs for these DOFs are limited to a minimum frequency of \SI{0.1}{\hertz} and a maximum frequency of \SI{50}{\hertz}. For roll and pitch rotation, the aVDS runs a direct angle cueing with an upper frequency limit of \SI{10}{\hertz}. Detailed MCA gain and filter settings are displayed in \autoref{tab_MCA} in the Appendix. For the variation of motion platform frequency content, the cut-off frequency of the low-pass filter is lowered to \SI{10}{Hz} for all DOFs resulting in an amplitude reduction of \SI{15}{\decibel} in the frequency range between $10$ and \SI{50}{\hertz} across all DOFs.

\subsubsection*{Road and tire model}
The reference road consists of sections with different surface conditions resulting in an overall degree of roughness of class C based on a one-track evaluation between the wavelengths of \SI{10}{\milli\metre} and \SI{10}{\kilo\metre} according to \cite{iso8608surface}. The composition of these sections is as follows: $19.9\%$ are of class A, $24.7\%$ of class B and $55.4\%$ of class C. In the simulator, these textures are represented by a horizontal \SI{10}{\milli\metre} grid with \SI{1}{\milli\metre} vertical resolution that was modeled from LiDAR data from the reference road. The tire is modeled as a steady-state slip model describing nonlinear slip forces and moments implemented as a Magic Formula 5.2 \cite{pacejka2005tire} parameter set that was parametrized using flat-track dyno measurements and validated using both open- and closed-loop measurements of the reference vehicle. The relaxation behavior is represented by a linear model using empirical relations for the relaxation lengths. The contact patch is represented using a set of unweighted contact points that are calculated via the cylindrical surface of the nominal tire radius and width that intersects with the road surface model. This intersection model returns the averaged contact patch center position and normal vector to the tire model. This simplification does have a significant effect on the resulting low-pass behavior of the representation of tire-road contact but was chosen since it represents a typical tire model for closed-loop evaluation of vehicle dynamics.

\input{fig_LongCtrlPerfo.tex}

\subsubsection*{Vehicle model}
The vehicle is represented by a two-track multibody model in IPG Carmaker. Rigid bodies represent chassis components and wheels. Elastokinematic effects of the suspension assembly between wheel carriers and the steering rack are represented by lookup tables that were parametrized via an elastic ADAMS multibody model and validated on a kinematics and compliance test rig. These tables also represent the changes in wheel orientation due to suspension kinematics depending on the steering rack position, vertical wheel deflection and wheel carrier forces. The steady-state behavior and transient response of the vehicle model were validated in accordance with \cite{iso22140validation,iso19634validation}.

\subsubsection*{Longitudinal control}
One key component of the chosen back-to-back study design was to ensure that the driver is experiencing the reference track at an identical velocity in all compared variants to achieve comparability of objective data in the frequency range and in doing so to validate the subjective comparison. For this purpose, each driver's velocity profile was recorded during the drive in the reference vehicle and was transformed into a coordinate-based velocity map during the return to the IFM, i.e. just before the driving simulator segment of the study. The resulting velocity map was then used to control the longitudinal vehicle motion during all simulator variants depending on the virtual vehicle's position on the track to match the road excitations during the virtual drives with the reference lap. Reference velocity tracking was achieved by an acceleration controller utilizing an MPC for the setpoint generation with phase compensation through velocity-dependent preview along the velocity map. This results in a Root Mean Square Error (RMSE) of below \SI{1}{\km/\hour} over the entire evaluation drive. \autoref{fig_LongCtrlPerfo} shows the resulting control performance relative to the corresponding vehicle speed.

\subsubsection*{Algorithm for rack force frequency augmentation}
To achieve the variations in SWT frequency content, we developed a real-time algorithm for deterministic augmentation of rack forces with measured high-frequency content that is specific to the individual trajectory and velocity profile during the reference drive of each participant. This algorithm for Rack Force Frequency Augmentation (RFFA) utilizes the rack force measurements taken during the reference drive to generate a rack force frequency map via a spatial arrangement of short-time fourier transform (STFT) coefficients along the reference path. This data analysis was performed during the drive back to the IFM to ensure a continuous back-to-back procedure.

\input{fig_PSD_FRack_AllVars.tex}

During model execution in the driving simulator, the algorithm reads the set of STFT coefficients corresponding to the current position on the track and applies the bandpassed difference of rack forces calculated through the vehicle and tire model and the inverse STFT coefficients as an additional force containing an equivalent of the frequency components between $10$ and \SI{30}{\hertz} that were measured during the reference drive. The lower bandpass limit of \SI{10}{\hertz} was chosen through subjective tuning by experts during a pilot study to ensure that low-frequency augmentation of SWT is not perceptible as an external steering intervention. Similar to the longitudinal controller, the RFFA makes use of velocity-dependent preview along the frequency map to compensate for the lumped system delay of rack force controller, signal latency, actuator dynamics and bandpass-phase. Additionally, a feed-forward gain based on the stationary frequency gain of the inverse transfer function of to HiL-actuator is applied to the rack force calculated by the RFFA to compensate for amplitude loss with increasing frequency. \autoref{fig_PSD_FRack_AllVars} illustrates the resulting closed-loop performance over the entire evaluation drive. \autoref{fig_AlgoPerformance} shows the augmentation of distinct spatial features as well as vehicle state-dependent dominant frequencies in variants with active RFFA.  

\input{fig_AlgoPerformance.tex}

\subsection{Questionnaire}
Post-drive evaluation of each simulator variant is split into two questionnaires. Both contain the same evaluation items but utilize different rating scales. In the reference vehicle, only the first questionnaire is being answered. It utilizes the established automotive assessment index (BI) described in \cite{heissing2002subjektive,harrer2007characterisation} to assess the participants' evaluation of system performance by means of a rating value between one and ten. In this study, a tendency indicating the direction of deviation from the optimum value, e.g. 'too low/high' is additionally stored in the sign of the BI. The second questionnaire utilizes a seven-point Likert scale to compare the steering feedback with the reference drive by means of a rating between significantly lower' and 'significantly higher' with the optimum value being 'identical'. Both rating scales are represented by a continuous scale with color-coded boxes in increments of $1$. User inputs are made via a tablet touchscreen, the minimum increment size is logged as $0.5$. The two rating scales are depicted in \autoref{fig_MXeval} in the Appendix. The aim of this separation is to isolate the subjective perception of differences in simulator validity from system performance and personal preference. A similar approach was chosen in previous studies \cite{zschocke2008links,zschocke2009beitrag,decker2009beurteilung}.

There are comprehensive works on established subjective criteria for the subjective assessment of steering feel \cite{riedel1997subjektive,harrer2007characterisation,decker2009beurteilung,rothhamel2011method,gomez2015findings}. In addition, numerous OEMs and steering system developers use their own evaluation catalogs, which are not published. The questionnaire used in this study is based on the published findings from the presented works, years of experience with steering feel development on the aVDS and the results from two pilot studies. Due to the specificity of the investigations made in the scope of this study and the heterogenous composition of participants regarding user level, it was necessary to make some adjustments to the scope of the evaluation catalogue. Only evaluation criteria related to the steering feedback behavior were chosen, there was no evaluation of vehicle guidance behavior. Additionally, two evaluation items that are not typically evaluated were added to the catalogue, both of which are aimed at a closer investigation of the criterion typically referred to as 'Road contact'. The additional items are 'Low-frequency road feedback (bumps and isolated events)' and 'High-frequency road feedback (vibrations)'. Since the majority of works in this field differentiate between beneficial and disturbing steering wheel feedback but no uniform delimitation of respective frequency ranges has been established, these additional questionnaire criteria aim at separating the subjective evaluation of these influences on driver perception. Of 30 participants, 28 answered the German version of the questionnaire, the remaining two used the English version. Both versions of the catalogue are presented in \autoref{tab_Questionnaire} in the Appendix.
In both pilot studies and related development work on the aVDS, a considerable number of participants expressed dissatisfaction regarding the overall level of road feedback through the steering system in their general feedback on system performance. Therefore, at any point during the evaluation drive and in the post-drive questionnaire, drivers were able to record personalized comments either by verbal feedback to the study operator or via text input on the evaluation tablet, in addition to the criteria from the questionnaire described above. As further explained in the discussion of limitations in chapter \ref{chap_Limitations}, this was one approach to ensure that drivers were able to express their personal preference for either of the systems or general dissatisfaction with system behavior across all variants without affecting the separation of variants and the detection of differences between the simulator variants and the reference. Additionally, after successful study conduct, the participants were asked which simulator variant they would prefer to use for an extended simulator drive based on their overall realism.

\input{tab_variants.tex}

\section{Results}
The subjective data obtained from the described questionnaires are well-interpretable for humans but have some properties that have a negative impact on the effectiveness of statistical analyses. For this reason, the evaluation of subjective data is divided into an initial examination of the distribution of raw subjective data to gain an understanding of the general result tendencies and a subsequent statistical analysis after a transformation into a more suitable format. In general, the evaluation is divided into the questionnaire for recording the subjective assessment of system performance using the BI and the comparison of the realism of the representation in the driving simulator using a Likert scale. The composition of the variants is summarized in \autoref{tab_variants}.

\subsection{Distribution of raw subjective data}
The choice of evaluation scales made in this study allows different approaches to the interpretation of the obtained subjective data. An initial investigation of absolute values of BI ratings without consideration of the direction expressed through their sign reveals a distribution of results with clear tendencies regarding both general rank as well as effect direction of the differences between the simulator variants. V2 is ranked worst in all evaluated characteristics, ratings of variants V1 and V3 show only small differences. Ratings of stationary characteristics show smaller differences than those related to Road contact. The reference vehicle outperforms all simulator variants in all characteristics related to Road contact. Simulator variant V3 performs better than V1, V2 performs worst. \autoref{fig_Spider_subFigures} shows these results in comparison to the reference vehicle. As expected, the result distribution is partly skewed with central tendencies closer to the ideal value than to the theoretical center of the scale. The center of result distributions is particularly close to the range of expected values for the stationary criteria and shows considerably lower values for the criteria related to Road contact. Mean values representing these central tendencies and variances, minimum and maximum values representing the dispersion of data are shown per variant in table \autoref{tab_MeanVar_Vehicle} in the Appendix. 

The result distribution exhibits considerable heterogeneity in terms of variance and normality through the individually varying utilization of the evaluation scale. Generally, some variance in scale usage across different users is to be expected, regardless of user level. Each subject uses a different distribution on the BI scale, both with regard to the center of the distribution and the spread of minimum and maximum values. \autoref{tab_MeanVar_UserLevel} in the Appendix shows the dependency of variance on the user level. Although all user groups exhibit the described variance in scale usage to some extent, overall result variances are considerably lower for higher user levels. This circumstance must be taken into account both from the point of usability by means of statistical analysis and with regard to the final interpretation of the results.

With the exception of the criterion Road contact, all subjective ratings from the first questionnaire are obtained using a bilinear scale. Without any adjustments to the raw data, this results in a partially non-continuous, non-normal data distribution due to the shift of the central tendency from the ideal value of 10. Therefore, the sign-weighted BI is instead converted into its sign-weighted deviation from the ideal BI rating of 10 preserving the directional intent of the information contained within the sign. \autoref{eq_inversion} shows the mathematical implementation of the described transformation.

\begin{equation}
  \label{eq_inversion}
  BI_{inv} = \sign(BI) \times (10 - |BI|)
\end{equation}

This results in a scale with its center at zero, a positive deviation represents an item rated as 'too high/strong/steep' while a negative deviation represents an item rated as 'too low/weak/flat'. While this transformation preserves the result distribution observed in the raw data and maintains interpretability, the consideration of the additional information contained in the sign results in a distinct two-peaked distribution of subjective values. This is to be expected for the evaluation of any variant that shows a deviation from the users' subjective ideal due to the definition of the BI. Therefore, non-normality of subjective data needs to be considered in the following statistical analysis.   

\input{fig_Spider_subFigures.tex}

\subsection{Data preparation for statistical analysis}
To mitigate the effect of different scale usage through individual users or user groups, the subjective ratings using the BI scale were normalized using a z-transform. The z-transform returns a $z$-score for each subjective rating such that the resulting distribution is centered to have a mean value of zero and scaled to have a standard deviation of 1. 

\newpage

\begin{equation}
  \label{eq_zTransform}
  z = \frac{(x - \overline{X})}{S}
\end{equation}

wherein $x$ is the current value of the input sample, $\overline{X}$ is its mean value and $S$ represents its standard deviation. 

The z-transformation, as shown in \autoref{eq_zTransform}, is performed across all criteria for each user. This way, by normalizing each user's responses relative to their own response distribution, the differences in central tendencies between evaluation criteria can be preserved while enabling the comparison of subjective ratings obtained from subjects of all user levels. An interpretation of transformed values is however more difficult. Due to the shift of the center of the distribution, the sign of the response no longer represents the direction of its deviation from the ideal value. Additionally, due to their scaling, the absolute values of z-scores allow no direct reference to the BI scale. As shown in previous works, if the primary goal of data preparation is different, such as a normalization of the result distribution \cite{harrer2007characterisation} or interpretability of central tendencies without further statistical analysis \cite{gomez2015findings}, a different transformation can be beneficial. In the present case, the primary goal is the homogenization of the variance distribution between drivers without the loss of resolution within a set of evaluations, which can be achieved through the z-transform, as shown in \cite{data2002objective,zschocke2008links}.

The subjective data obtained from the second questionnaire are already in a Likert scale format with a minimum value of $-3$ and a maximum value of $3$. The sign convention is identical to the one used in the BI questionnaire after transformation with \autoref{eq_inversion}. Hence, for interpretation of results from the back-to-back comparison as well as comparison with the BI results, this format will be chosen in the following data analysis and discussion of results. The Likert data from the second questionnaire show good variance homogeneity across all user levels indicating that scale usage is considerably more similar between different users and user levels but are non-normal distributed. Hence, these data are normalized using the z-transform as well. All data preparation was performed using MATLAB 2023a.

\clearpage

\section{Statistical analysis}
An analysis of variances (ANOVA) was performed on the subjective ratings collected in the driving simulator to compare the significance of each variant's effects on subjective steering feedback evaluation with a significance level of $.05$. All data were tested for normality using Shapiro-Wilk method. Additionally, a test of homogeneity of variances using Levene's method was performed. For subjective data following a normal distribution, a conventional parametric repeated measures ANOVA was performed. Pairwise comparison of variants was performed using Tukey’s HSD Test. For the large part of subjective data that deviated significantly from a normal distribution or showed significant violations of homogeneity of variance, a non-parametric ANOVA using Friedman's method was performed. Pairwise comparison of variants was performed using jamovi's Durbin-Conover method \cite{pohlert2014pairwise} (PMCMR). The analysis of comparison ratings was performed across subjective data from simulator variants and analysis of BI-ratings was performed across the whole dataset including the evaluations from the reference vehicle. All statistical analysis was performed in jamovi version 2.3.28.

\input{tab_pairwise_BI.tex}

\subsection{BI ratings of simulator variants and reference vehicle}
A repeated measures ANOVA revealed that there was a highly significant difference in subjective ratings of the parameter Road contact \anovaPL{3}{72}{19.404}{}. A Friedman test revealed that there was a highly significant difference in subjective ratings of the parameters Low-frequency road contact \ChiSquarePL{3}{18.8}{} and High-frequency road contact \ChiSquarePL{3}{32.4}{}. Pairwise comparisons of these parameters are presented in \autoref{tab_pairwise_BI}. All variants including the reference were ranked with negative values, i.e. too low/weak/flat across all displayed parameters. Therefore, a positive mean value difference indicates that a variant was rated better, regardless of absolute values. All other questions did not show significant differences between either the reference vehicle and simulator variants or between simulator variants.

\subsection{Comparison ratings of simulator variants}
A Friedman test revealed that there was a highly significant difference in subjective ratings of the parameters Road contact \ChiSquarePL{2}{16.3}{} and High-frequency road contact \ChiSquarePL{2}{23.6}{} between at least two simulator variants. Pairwise comparisons of these parameters are presented in \autoref{tab_pairwise_Comp}.

\input{tab_pairwise_Comp.tex}

\subsection{Differences between user levels}
An analysis of the subjective evaluation data of all evaluated variants separated by user level shows that in the first questionnaire (BI) the entire sample shows significant mean value differences in the same criteria as the subgroups of experts and non-experts. The results do not show significant mean value differences in any of the evaluation items related to stationary steering feedback.

An analysis of the subjective evaluation data of simulator variants separated by user level concludes that in the second questionnaire (Comparison) the entire subject sample shows significant mean value differences in the same criteria as the subgroups of experts and non-experts.

Overall, criteria that showed significant or marginal mean value differences among non-experts, showed significant mean value differences with stronger significance among experts.

\subsection{Effect of simulator variations on subjective evaluations}
T-tests for both modifications that were in effect between the simulator variants were performed for all criteria that showed significant differences in the ANOVA. Since the modifications can be treated as independent factors applied to the simulator variants and each subject provided evaluations for combinations of both modifications, an independent sample t-test was chosen. Data following a normal distribution were analyzed with Student's t-test when homogeneity of variances was given and Welch's t-test else. Non-normal distributed data was analyzed through a Mann-Whitney U-test when homogeneity of variances was given and Yuen's robust t-test else.

In the first questionnaire, Welch's t-test showed that the RFFA variation had a highly significant effect on the subjective evaluation of Road contact (\studentTPL{79}{-3.818}{}), while a Mann-Whitney U-test revealed a highly significant effect on Low-frequency road contact (\mannWhitney{479}{0.012}) and a robust t-test revealed a highly significant effect on High-frequency road contact (\yuenT{22.9}{2.92}{0.008}). 

A Mann-Whitney U-test showed that the MCA variation had a highly significant effect on subjective evaluation of both Road contact (\mannWhitneyPL{399}{}) and Low-frequency road contact (\mannWhitney{518}{0.035}).

In the second questionnaire, a Mann-Whitney U-test showed that the RFFA variation had a highly significant effect on the subjective evaluation of comparison ratings of both Road contact (\mannWhitneyPL{364}{}) and Low-frequency road contact (\mannWhitneyPL{284}{}). 

A Mann-Whitney U-test showed that the MCA variation had a highly significant effect on the subjective evaluation of comparison ratings of both Road contact (\mannWhitney{426}{0.002}) and High-frequency road contact (\mannWhitney{424}{0.002}). 

Across all users, Levene's test revealed that the RFFA variation resulted in a significant variance inhomogeneity for the criteria SWT off-center with \anova{1}{79}{4.47}{0.0038} and High-frequency road contact with \anova{1}{79}{4.516}{0.037}. In the subgroup of non-experts, this effect was even more pronounced with the items SWT on-center \anova{1}{55}{5.522}{0.022}, SWT off-center \anova{1}{55}{5.266}{0.026}, SWT gradient (\anova{1}{55}{4.008}{0.050}) and High-frequency road contact \anova{1}{55}{6.652}{0.013} showing significant variance inhomogeneity in Levene's test while for expert results this was only the case for the criterion Steering wheel torque off-center with \anova{1}{22}{4.768}{0.040}. 

\section{Discussion}
The statistical analysis of z-transformed subjective data revealed that the conclusions that can be drawn based on the ranking of raw data of the evaluated variants from both questionnaires are consistent with each other and stand up to statistical verification. Overall, variant V3 was rated best, V2 performed the worst across the board. Statistical analysis of the BI-questionnaire results showed considerably lower significance in mean value differences between the reference vehicle and V3 compared to the other simulator variants across all user groups. These findings coincide with the results of the ranking based on the final post-study evaluation regarding overall realism of the steering feedback whereby 21 participants chose V3, four participants chose V1 and only two participants chose V2. 

One potential limitation to this statement is the fact that some drivers reported an overall unsatisfactory level of road feedback in the reference vehicle which might influence subjective evaluations in the direction of a general preference of variants providing more feedback. The observation of a generally insufficient level of road feedback is consistent with the results of existing studies that utilized recordings from steering wheel excitations in a lab testing environment \cite{giacomin2005study,berber2009evaluation} although transfer of these results might be limited due to the substantial differences in driver perception between the different environments as well as fundamentally different steering system properties and representation fidelities under investigation. To mitigate this effect, the questionnaire was specifically separating the evaluation of quality or system performance via the BI which can be strongly influenced by personal liking from the evaluation of perceived differences between the reference and the driving simulator. Additionally, in the briefing before the study, special emphasis was placed on conveying that the aim of the driving simulator study was not to tune or improve the presented steering feel in the vehicle, but to reproduce the reference system as realistically as possible. The comparison results from the second questionnaire are consistent with the results from the first questionnaire regarding items related to stationary steering feedback but show a difference regarding the influence of the simulator modifications on the individual items related to Road contact. Additionally, results from the second questionnaire show higher consistency regarding the variants that were perceived as significantly different from each other with both variants V1 and V2 showing significant differences from V2 but no significant differences between each other. Hence, it can be concluded that in a back-to-back study design as presented, the second questionnaire provides additional information beyond the scope of the results from the traditional steering feel questionnaire. Furthermore, these findings suggest that the differences in High-frequency road contact are more important for the perceived differences in Road contact and an isolation of these criteria is more effective when participants are asked to evaluate perceived differences rather than system performance.

Overall, subjective data showed statistically significant differences only between the three characteristics that are related to Road contact. Thus, it can be concluded that the characteristics of the simulation environment that are related to the stationary steering properties were represented with sufficient realism to have no significant influence in the presented study design. Therefore, the simulation environment can be considered behaviorally validated for the presented use case of isolating the subjective perception of Road contact through steering feedback.
The ANOVA results revealed a significant difference between the subjective evaluations of all three simulator variants in at least one characteristic related to Road contact with V3 showing the least significant differences, again coinciding with the overall rank of the variants. This means that as initially hypothesized, significant differences in subjective perception of steering feedback in dynamic driving simulators can be found when no significant deviation of objective steering feel parameters or subjective evaluation of static steering feedback exists. More specifically, this shows that established evaluation characteristics regarding stationary steering feedback forces and dynamic rotational steering feedback forces in the frequency range up to \SI{30}{\hertz} do not cover all relevant factors for the subjective evaluation of steering feedback in dynamic driving simulators. 

T-tests for both modifications that were made between simulator variants revealed that there was a highly significant effect of both vehicle motion frequency content between $10$ and \SI{50}{\hertz} and rack force frequency content between $10$ and \SI{30}{\hertz} on the subjective evaluation of Road contact and Low-frequency road contact in all subjects. Effect size was considerably higher for the effect on Road contact, variants with RFFA scored higher than without RFFA in both questionnaires. Rack force frequency content additionally showed a highly significant effect on the subjective evaluation of High-frequency road contact. This shows that the questionnaire design allowed for isolation of the effects introduced by the steering wheel vibrations while the established criterium Road contact is not suited to differentiate between the contribution of vehicle and steering wheel vibration towards the subjective perception of steering feedback. Additionally, despite the heterogeneity of participants' skill and experience levels, an evaluation of even more finely separated criteria than typically used by steering experts is possible with an adjusted study design and in-depth technical briefing of participants. With regards to participants' skill levels, the results generally show that experts were able to identify differences in more steering feedback characteristics and between more variants than non-experts. Non-expert ratings did not contradict those made by experts, i.e. all criteria that showed significant or marginal mean value differences among non-experts, showed significant mean value differences with stronger significance among experts as well. For this reason, the participation of non-experts resulted in an increase in statistical power through increased sample size without limiting the differentiability of the entire sample. However, it can be assumed that this effect can not be generalized and depends on the composition of the test subject collective and several factors in study design. 

\input{fig_Box_UserLevels.tex}

Although not statistically significant across the entire dataset, raw and transformed results showed differences in the evaluation of stationary steering feedback criteria in some cases. This was particularly noticeable for the criterion Steering wheel torque off-center which even exhibits a statistically significant mean value difference in BI-ratings with \ChiSquare{2}{6.15}{0.046} when the simulator variants are analyzed separately, albeit with an insufficient effect size of $\eta^2 = 0.034$. This effect does not coincide with the findings from the second questionnaire and does not carry over to the findings from the entire dataset. An analysis of subgroup ratings shows that this mean value difference can only be observed among non-experts suggesting that this effect is strongly influenced by a between-subjects effect caused by the parameter User level. Across the entire dataset, any mean value differences between subgroups were insignificant, as reported above. Furthermore, reducing the number of observations for the isolated analysis of simulator variants via a Friedman test has the effect of lowering statistical power with the result of an increased risk of type one errors. Hence, it can be concluded that, if at all statistically relevant, the mean value difference of this criterion can most likely be attributed to a limited understanding of this criterion among non-experts rather than a perceived difference in stationary steering wheel torque. However, this observation highlights the influence of simulator variations on the results of the stationary characteristics. The rating distribution in \autoref{fig_Box_UserLevels} shows that although no statistically significant mean value differences can be reported, an influence on the data quality is noticeable. The evaluations from the subgroup of experts show considerably decreased variances for both simulator variants with active RFFA. Although sample size within subgroups is insufficient for a conclusion with statistical power, this observation supports the findings from the variance analysis. As described above, results from Levene's test show a strong influence of steering rack force frequency content on the evaluation variance of all stationary steering feedback criteria. It can therefore be assumed that the evaluation quality of steering feedback in a dynamic driving simulator can be affected by the SWT frequency content.

\section{Limitations}
\label{chap_Limitations}
One potential limitation of this study is the limited sample size and the strong imbalance in the gender distribution of the participants. This is particularly relevant for the investigation of effects within subgroups of participants and has a negative impact on the significance for the overall population. However, due to the high specificity of the research questions to be investigated, study design was tailored towards the development processes typically involving driving simulators for subjective evaluation of vehicle dynamics. Hence, the importance of the evaluation through experienced simulator drivers that are capable of performing subjective evaluations without being overwhelmed by an unfamiliar environment was crucial. Additionally, since this study investigated the differences between simulator variants with very similar characteristics, the participation of system experts who perform similarly specific steering feel evaluations on a regular basis was substantial to the study design. Since the recruitment of a relevant number of such specifically trained experts is particularly difficult, an even gender distribution and generally increased sample size could not be achieved in the scope of this research project. This makes an evaluation of gender-specific differences not possible which could provide additional insights regarding the driver specific adaptation of steering feedback. While this does not fall within the scope of this study, it might become more relevant with the increasing prevalence of SbW systems and should therefore be explored in future research. 

Secondly, there is a conceptual limitation regarding the comparability of closed-loop experimental data which is caused by the limited repeatability of driver inputs resulting in deviations of the vehicle reaction between consecutive evaluation drives. In the case of this study, they arise from the fact that the reference drives were performed without longitudinal control engaged whereas in the driving simulator, longitudinal control was performed by a controller model to precisely match the recorded velocity from the reference drive. While this does enable the comparison of objective data and ensures that drivers experience the reference track at the same conditions with every evaluated steering feel variant, it has the negative effect of a change in driver load between the reference drive and the virtual evaluation drives. Beyond physical reasons for the importance of accurate representation of excitation frequencies, the choice of driving speed has been proven to have a significant effect on the subjective perception of steering feedback torque in driving simulators \cite{bertollini1999applying,neukum2010untersuchung}. Since the reference drive takes place on public roads, it is not safely possible to let the reference vehicle take care of longitudinal control over the whole evaluation drive. Furthermore, several simulator studies have shown that drivers can show significant differences in driving behavior including choice of velocity between real drives and driving simulators \cite{fischer2012evaluation,van2018relation,bellem2018comfort}. A benefit of this design is the fact that drivers typically choose a velocity more carefully in a real vehicle which results in an evaluation environment with a lower probability of overwhelming the driver. Additionally, since the participants are driving the same reference track four times in a row, free longitudinal control through the driver would introduce the possibility of training effects on velocity choice. Due to the focus of the study on the influence of road-induced steering wheel and motion platform vibrations, the consistent replication of the velocity and consequently frequency content experienced in the reference vehicle is crucial and is therefore prioritized over the replication of driver workload in the driving simulator. If this resulted in altered perception thresholds in the virtual evaluation drives these are very likely to be consistent across the entire sample since every driver was only evaluating all variants with their own velocity profile and there were no changes between simulator variants. Ultimately, this design allows an interpretation of subjective data without additional correlation with objective data between simulator variants since all participants experienced the same level of physical validity. 

\section{Conclusion} 
In conclusion, a controlled back-to-back study was conducted using a reference vehicle and three variants of its virtual representation in a dynamic high-fidelity driving simulator to assess the influence of steering wheel feedback torque frequency content between $10$ and \SI{30}{\hertz} and vehicle motion frequency content between $10$ and \SI{50}{\hertz} on the subjective evaluation of steering feedback with a focus on parameters related to Road contact in a dynamic driving simulator. Participants evaluated three simulator variants with a combination of two different steering wheel feedback frequency contents and two different vehicle motion frequency contents in direct comparison after a drive in the reference vehicle.

Results show that while with a state-of-the-art steering feedback representation, physical validity in established steering feel evaluation characteristics can be achieved to a sufficient degree, this does not result in an equivalent level of behavioral validity regarding the subjective evaluation of steering feedback in a dynamic driving simulator. A statistical analysis found significant differences from the subjective evaluation of the reference vehicle in all criteria related to Road contact. Participants of all user groups could identify effects on all relevant characteristics of steering feedback. It was demonstrated that both rack force frequency content and vehicle motion frequency content in the investigated ranges can influence the data quality of the evaluation of stationary steering feedback characteristics. These effects were shown to have a considerable between subjects effect depending on User level. Lastly, all tested variants showed significant mean value differences from the reference vehicle in the evaluation of both Road contact and High-frequency road contact indicating that more research of relevant influence factors is necessary to achieve behavioral validity in all steering feedback characteristics in a dynamic driving simulator.

Future research should therefore investigate the effect of system excitations that were not within the scope of the presented study such as non-road-induced steering wheel and vehicle body excitations. Additionally, an investigation of the effects of higher excitation frequencies via both the tactile and acoustic channel on driver perception of steering feedback in dynamic driving simulators is recommended. 

\section*{Contribution}
Maximilian Böhle initiated the idea of this paper and designed and implemented the presented subject study. Steffen Müller and Bernhard Schick made essential contributions to the conception of the research project and the study design and revised the paper critically.

\section*{Acknowledgments}
Special thanks go to Johann Haselberger of Kempten University of Applied Sciences for the knowledgeable technical discussions and constructive ideas regarding the statistical analysis of the study results.

\bibliographystyle{IEEEtran}
\bibliography{IEEEabrv,2024_IEEE_T-IV.bib}
\FloatBarrier
\newpage

\begin{IEEEbiography}[{\includegraphics[width=1in,height=1.25in,clip,keepaspectratio]{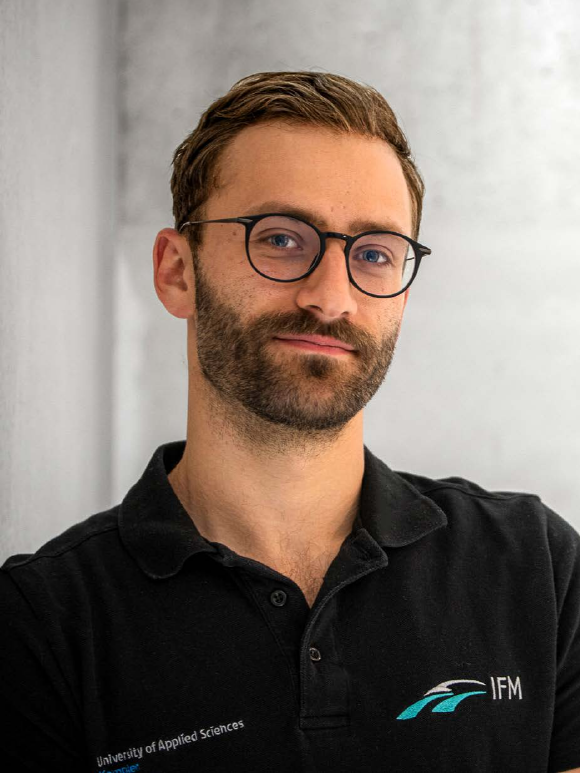}}]{Maximilian Böhle}
  received the B.Sc. and M.Sc. degrees in Automotive Engineering at Munich University of Applied Sciences, Germany in 2016 and 2019. He is currently working towards his doctoral degree in Automotive Engineering at the Faculty of Mechanical Engineering and Transport Systems of the Technical University of Berlin, Germany. Since 2020, he has been working as a research assistant at the Institute for Driver Assistance and Connected Mobility at Kempten University of Applied Sciences, Germany. His research focuses on Steering feel and Vehicle dynamics for Driving simulators.
\end{IEEEbiography}
\vspace{-7cm}
\begin{IEEEbiography}[{\includegraphics[width=1in,height=1.25in,clip,keepaspectratio]{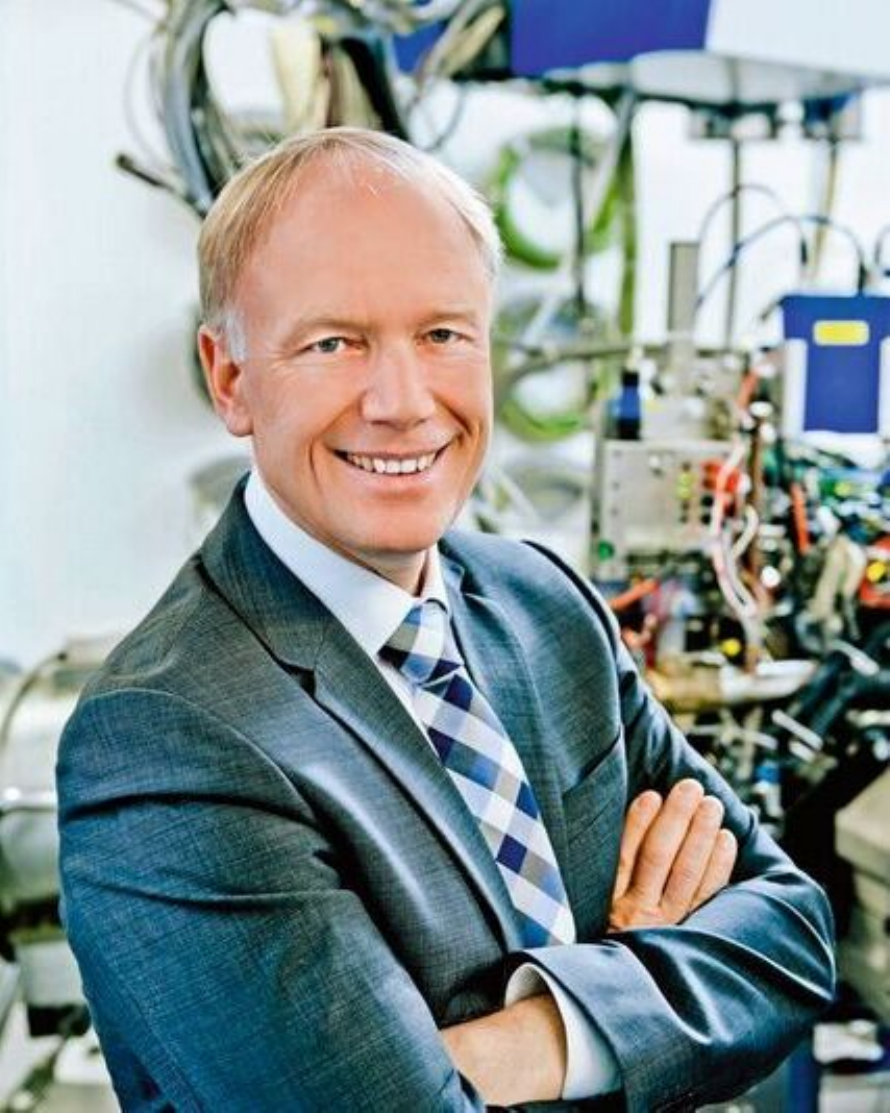}}]{Bernhard Schick}
  holds a degree in mechatronic engineering at the University of Applied Sciences Heilbronn. From 1994, whilst at TÜV SÜD, he built up his expertise in the field of vehicle dynamics and advanced driver assistance systems, in various positions up to a general manager. He joined IPG Automotive in 2007 as managing director, where he worked in the field of vehicle dynamics simulation. From 2014 he was responsible for calibration and virtual testing technologies as global business unit manager at AVL List, Graz. Since 2016, he has been a research professor at the University of Applied Sciences Kempten and the Head of the Institute for Driver Assistance Systems and Connected Mobility. His research focus is automated driving and vehicle dynamics.
\end{IEEEbiography}
\vspace{-7cm}
\begin{IEEEbiography}[{\includegraphics[width=1in,height=1.25in,clip,keepaspectratio]{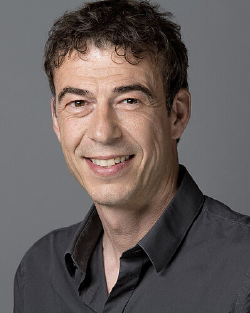}}]{Steffen Müller}
  received the Dipl.-Ing. degree in astronautics and aerospace engineering in 1993 and the Dr.-Ing. degree from the Technical University of Berlin in 1998. From 1998 to 2000, he was a project manager at the ABB Corporate Research Center, Heidelberg, Germany. He finished his post-doctoral research at the University of California, Berkeley, in 2001. From 2001 to 2008, he held leading positions at the BMW Research and Innovation Centre. From 2008 to 2013, he was the founder and leader of the Chair for Mechatronics in Engineering and Vehicle Technology, Technical University of Kaiserslautern, Germany. He is a University Professor and an Einstein Professor at the Technical University of Berlin, Germany. He is the Head of the Chair of Automotive Engineering, Faculty of Mechanical Engineering and Transport Systems.
\end{IEEEbiography}

\clearpage

\input{tab_GPower.tex}
\input{tab_UserLevels.tex}
\input{tab_MCA.tex}

\clearpage
\input{tab_MeanVar_Vehicle.tex}

\input{tab_MeanVar_UserLevel.tex}

\input{fig_MXeval.tex}
\input{tab_Questionnaire.tex}

\end{document}

%% file: ClassicKPIs.tex
\begin{table}[]
  \centering
  \caption{Objective validity of classic steering feel \\characteristics: Reference values marked with \textsuperscript{*}}
  \label{tab_KPIs}

  \begin{tabular}{llc}
    \toprule 
    \textbf{Characteristic}  & \textbf{Parameter}  & \textbf{Value} \\
    \midrule

    Steering wheel torque   & Steering stiffness at                   & 0.29 \textsuperscript{*}  \\
    level and gradient      & zero steer (\SI{}{\newton\metre\per\deg}) & 0.29 \textsuperscript{ } \\[0.2cm]
                            & Steering friction (\SI{}{\newton{}\metre})                  & 1.03 \textsuperscript{*} \\
                            &                                         & 0.99 \textsuperscript{ } \\[0.2cm]
    On-center precision     & Steering wheel angle                    & 7.95 \textsuperscript{*} \\
                            & hysteresis (\SI{}{\deg})                        & 7.76 \textsuperscript{ } \\[0.2cm]
    Road feedback           & Bandwidth of road induced               & 17.4 \textsuperscript{*} \\
                            & steering wheel torque (\SI{}{\hertz})              & 17.4 \textsuperscript{ } \\
    \bottomrule
  \end{tabular}
  \end{table}

%% file: fig_aVDS.tex
\begin{figure}[!t]
    \centering
    \includegraphics[width=3.5in]{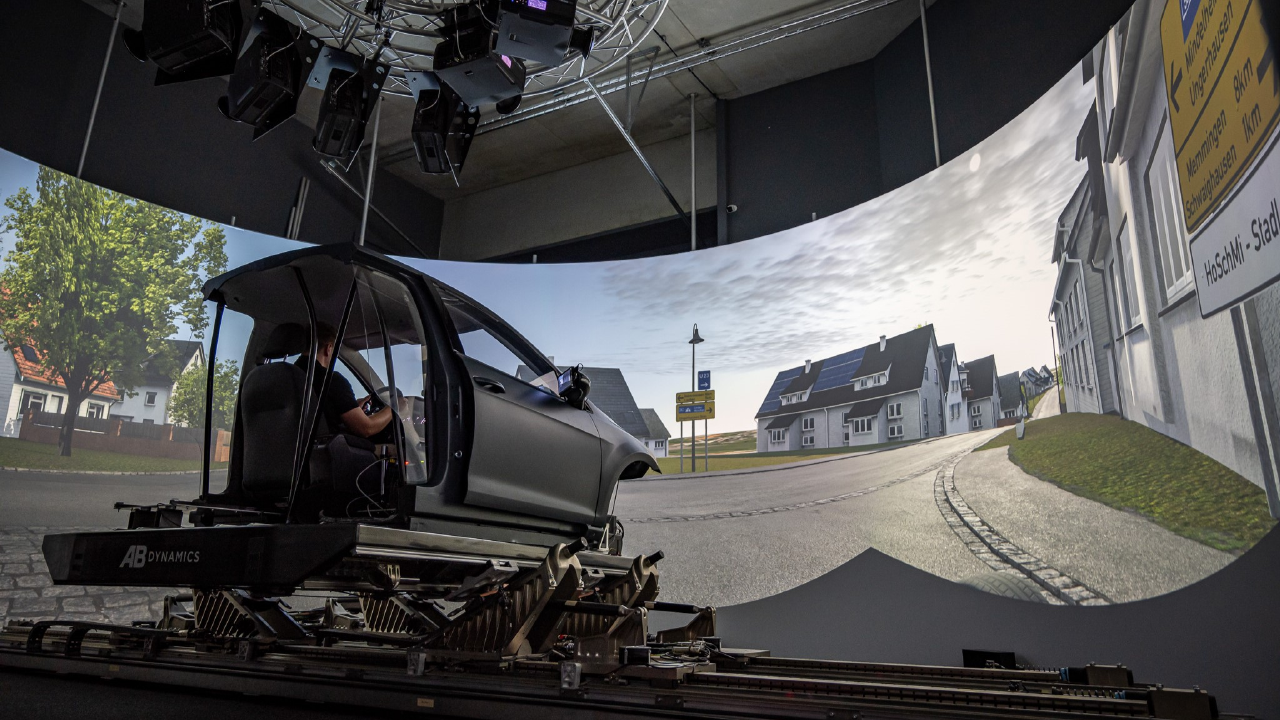}
    \caption{The Advanced Vehicle Driving Simulator
    (aVDS) at the IFM of the Kempten University of Applied Sciences}
    \label{fig_aVDS}
\end{figure}

%% file: AVDS_performance.tex
\begin{table}[]
\centering
\caption{Performance of the 6-DOF motion platform according to \cite{abd2024bandwidth}}
\label{tab_AVDS_performance}
\begin{tabular}{lll}
\toprule
\textbf{Degree of   Freedom} & \textbf{Effective travel} & \textbf{Bandwidth} \\ \midrule
Surge   (translation in X)   & $\pm$ \SI{540}{\milli\metre}     & $>$ \SI{15}{\hertz}           \\
Sway   (translation in Y)    & $\pm$ \SI{1250}{\milli\metre}     & $>$ \SI{35}{\hertz}           \\
Heave   (translation in Z)   & $\pm$ \SI{120}{\milli\metre}     & $>$ \SI{50}{\hertz}           \\
Roll (rotation   around X)   & $\pm$ \SI{8}{\degree}     & $>$ \SI{50}{\hertz}           \\
Pitch   (rotation around Y)  & $\pm$ \SI{9}{\degree}     & $>$ \SI{50}{\hertz}           \\
Yaw (rotation   around Z)    & $\pm$ \SI{30}{\degree}     & $>$ \SI{35}{\hertz}           \\ \bottomrule
\end{tabular}
\end{table}

%% file: fig_LongCtrlPerfo.tex
\begin{figure}[!t]
    \centering
    \includegraphics[width=3.5in]{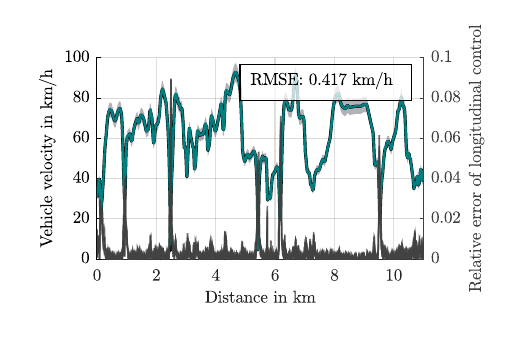}
    \caption{Performance of the longitudinal controller. The solid black line on the left axis represents the reference velocity profile, its 95 percent confidence interval is displayed in grey. The mint-colored lines represent the velocity profiles from all three simulator variants. The grey lines on the right axis represent the respective relative error. The displayed RMSE value is the mean over all three displayed simulator variants.}
    \label{fig_LongCtrlPerfo}
\end{figure}

%% file: fig_PSD_FRack_AllVars.tex
\begin{figure}[t]
    \centering
    \includegraphics[width=3.5in]{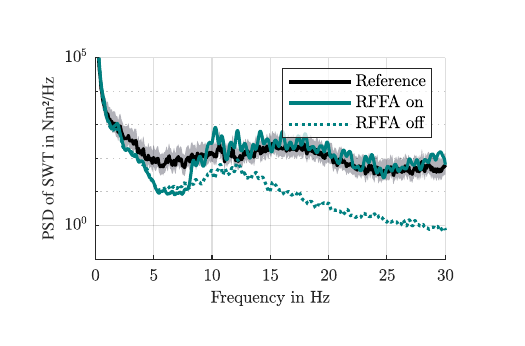}
    \caption{Comparison of power spectral densities of SWT during closed-loop driving. The black line represents the reference drive, its 95 percent confidence interval is displayed in grey. The mint-colored lines represent the simulator variants. The solid line represents the simulator variants with active RFFA while the dotted (lower) line represents the simulator variants without RFFA.}
    \label{fig_PSD_FRack_AllVars}
\end{figure}

%% file: fig_AlgoPerformance.tex
\begin{figure*}[t]
    \centering
    \subfloat[]{\includegraphics[width=2.33in]{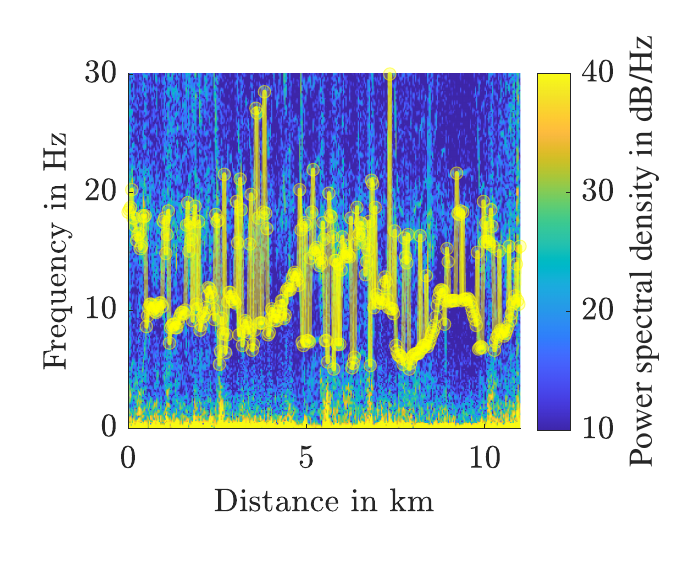}%
    \label{fig_AlgoPerformance_ref}}
    \hfil
    \subfloat[]{\includegraphics[width=2.33in]{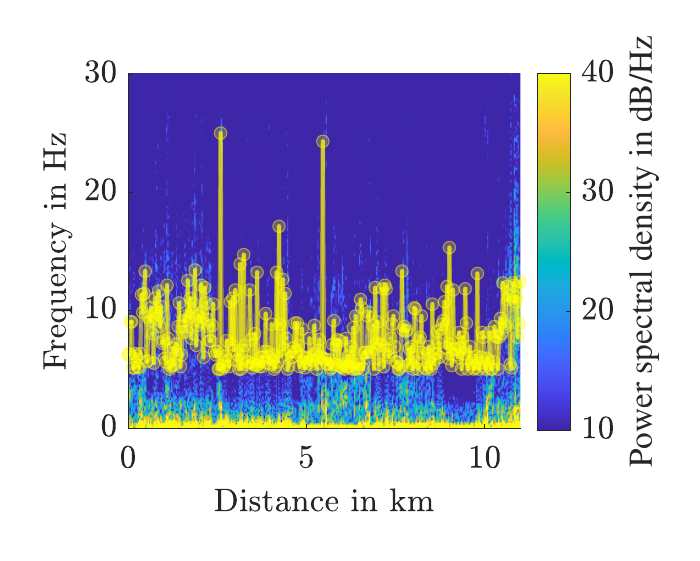}%
    \label{fig_AlgoPerformance_V2}}
    \hfil
    \subfloat[]{\includegraphics[width=2.33in]{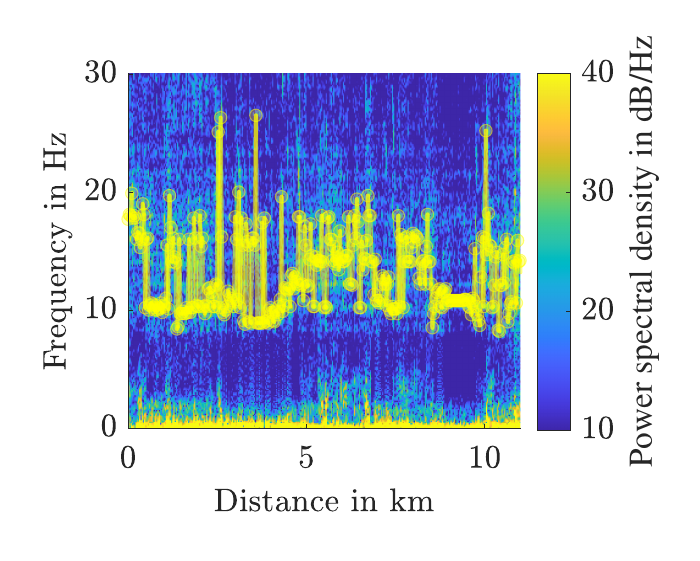}%
    \label{fig_AlgoPerformance_V3}}
    \caption{Spectrogram of measured rack forces vs. distance traveled along the reference path. \ref{fig_AlgoPerformance_ref} Reference vehicle, \ref{fig_AlgoPerformance_V2} aVDS with raw model output (V2) and \ref{fig_AlgoPerformance_V3} aVDS with RFAA (V1 and V3). The yellow scatters mark the dominant frequency at each STFT sample.}
\label{fig_AlgoPerformance}
\end{figure*}

%% file: tab_variants.tex
\begin{table}[htbp]
    \centering
    \caption{Simulator variants: V1 and V3 with active RFFA, V1 and V2 with full MCA low-pass cut-off frequency ($f_{LPF}$) of \SI{50}{\hertz}.}
      \begin{tabular}{ccc}
        \toprule
      \textbf{Variant} & \textbf{RFFA} & \textbf{MCA $f_{LPF}$} \\
      \midrule
      V1    & \cmark & \SI{50}{\hertz} \\
      V2    & \xmark & \SI{50}{\hertz} \\
      V3    & \cmark & \SI{10}{\hertz} \\
      \bottomrule
      \end{tabular}%
    \label{tab_variants}%
  \end{table}%

%% file: fig_Spider_subFigures.tex
\begin{figure*}[!t]
\centering
\subfloat[]{\includegraphics[width=2.33in]{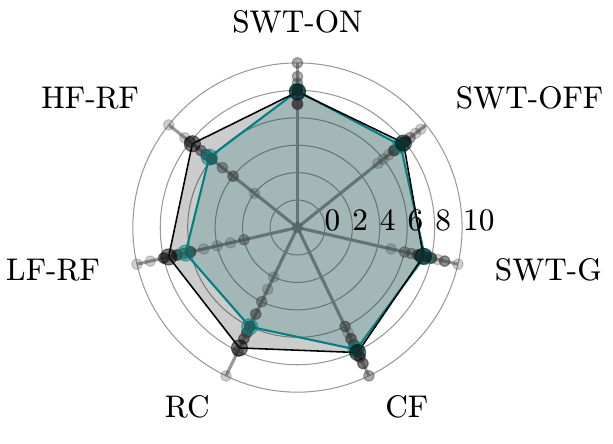}%
\label{fig_Spider_RoadContactVar139}}
\hfil
\subfloat[]{\includegraphics[width=2.33in]{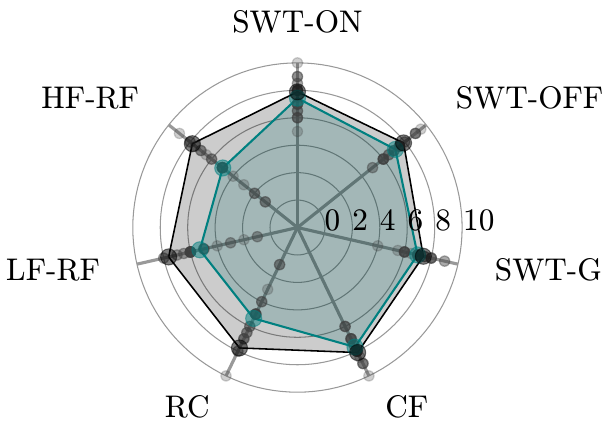}%
\label{fig_Spider_RoadContactVar239}}
\hfil
\subfloat[]{\includegraphics[width=2.33in]{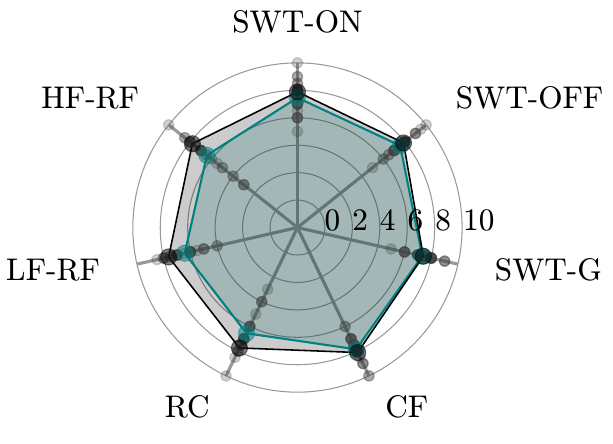}%
\label{fig_Spider_RoadContactVar339}}
\caption{Absolute values of BI ratings (higher value means better) for simulator variants V1 (\ref{fig_Spider_RoadContactVar139}), V2 (\ref{fig_Spider_RoadContactVar239}) and V3 (\ref{fig_Spider_RoadContactVar339}). The mint-colored lines represent the simulator variants while the black lines represent the reference vehicle. SWT-ON: 'SWT on-center', SWT-OFF: 'SWT off-center', SWT-G: 'SWT gradient', CF: 'Centre-feel', RC: 'Road contact', LF-RF: 'Low-frequency road feedback', HF-RF: 'High-frequency road feedback'}
\label{fig_Spider_subFigures}
\end{figure*}

%% file: tab_pairwise_BI.tex
\begin{table}[]
    \centering
    \caption{Pairwise comparison of BI ratings. The variant that performed better is shown in bold. Positive mean difference means B was rated higher/stronger/steeper, i.e. better than A.}
    \label{tab_pairwise_BI}
    \begin{tabular}{lllrrr}
    \toprule
    \multicolumn{3}{c}{\textbf{Comparison}} & \textbf{Mean difference}      & \textbf{p}    \\ 
    A                  & - & B          &                               &               \\ \midrule
    \multicolumn{5}{l}{\textit{Road contact} \textsuperscript{*}} \\ \midrule
    \textbf{Ref}       & - & V1                & $-0.724$                      & $< .001$   \\
                       & - & V2                & $-1.062$                      & $< .001$   \\          
                       & - & V3                & $-0.464$                      & $ 0.008$    \\
    \textbf{V1}        & - & V2                & $-0.338$                      & $ 0.051$    \\
    V1                 & - & \textbf{V3}                & $0.261$                       & $ 0.021$    \\          
    V2                 & - & \textbf{V3}                & $0.599$                       & $ 0.002$    \\ \midrule
    \multicolumn{5}{l}{\textit{Low-frequency road contact} \textsuperscript{**}} \\ \midrule
    \textbf{Ref}       & - & V1                & $-0.408$                      & $ 0.024$    \\
                       & - & V2                & $-0.724$                      & $< .001$   \\          
                       & - & V3                & $-0.170$                      & $ 0.229$    \\
    \textbf{V1}        & - & V2                & $-0.317$                      & $ 0.021$    \\
    V1                 & - & \textbf{V3}                & $0.238$                       & $ 0.282$    \\          
    V2                 & - & \textbf{V3}                & $0.554$                       & $< .001$   \\ \midrule
    \multicolumn{5}{l}{\textit{High-frequency road contact} \textsuperscript{**}} \\ \midrule
    \textbf{Ref}       & - & V1                & $-0.855$                      & $< .001$   \\
                       & - & V2                & $-1.527$                      & $< .001$   \\          
                       & - & V3                & $-1.015$                      & $< .001$   \\
    \textbf{V1}        & - & V2                & $-0.672$                      & $ 0.015$    \\
    V1                 & - & \textbf{V3}                & $-0.16$                       & $ 0.324$    \\          
    V2                 & - & \textbf{V3}                & $0.512$                       & $< .001$    \\\bottomrule
    \vspace{-0.2cm} \\
    \multicolumn{5}{l}{\textsuperscript{~ *}: Tukey's HSD} \\
    \multicolumn{5}{l}{\textsuperscript{**}: PMCMR} \\
    \end{tabular}
    \end{table}

%% file: tab_pairwise_Comp.tex
\begin{table}[]
    \centering
    \caption{Pairwise comparison of comparison ratings (PMCMR). The variant that performed better is shown in bold.}
    \label{tab_pairwise_Comp}
    \begin{tabular}{lllrrr}
    \toprule
    \multicolumn{3}{c}{\textbf{Comparison}} & \textbf{Mean difference}      & \textbf{p}    \\ 
    A                  & - & B          &                               &               \\ \midrule
    \multicolumn{5}{l}{\textit{Road contact}} \\ \midrule
    \textbf{V1}        & - & V2                         & $-0.333$                      & $ 0.004$    \\
    V1                 & - & \textbf{V3}                & $ 0.233$                      & $ 0.111$    \\          
    V2                 & - & \textbf{V3}                & $-0.566$                      & $< .001$    \\ \midrule
    \multicolumn{5}{l}{\textit{High-frequency road contact}} \\ \midrule
    \textbf{V1}        & - & V2                         & $ 0.555$                      & $< .001$    \\
    V1                 & - & \textbf{V3}                & $-0.183$                      & $ 0.184$    \\          
    V2                 & - & \textbf{V3}                & $-0.738$                      & $< .001$    \\\bottomrule
    \end{tabular}
    \end{table}

%% file: fig_Box_UserLevels.tex
\begin{figure*}[!t]
    \centering
    \subfloat[]{\includegraphics[width=3.5in]{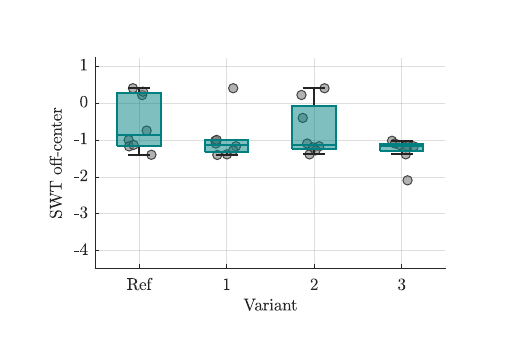}%
    \label{fig_Box_OnCenterExperts}}
    \hfil
    \subfloat[]{\includegraphics[width=3.5in]{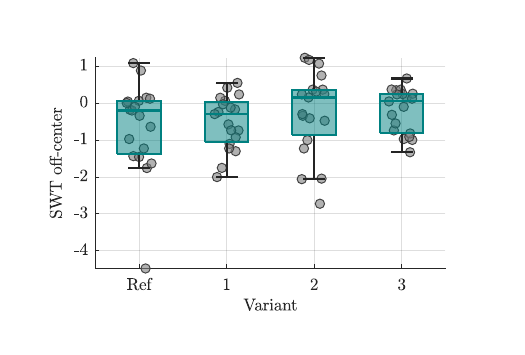}%
    \label{fig_Box_OnCenterNonExperts}}
    \caption{Comparison of z-transforms of subjective ratings for Steering wheel torque off-center between experts (a) and non-experts (b).}
    \label{fig_Box_UserLevels}
\end{figure*}

%% file: tab_GPower.tex
\begin{table*}[h!]
    \centering
    \caption{A priori power analysis for calculation of necessary \\sample size using G*Power \cite{faul2007gpower}.}
      \begin{tabular}{llll}
        \toprule
        \textit{ANOVA: Repeated measures, within factors} \\
        \midrule
      \textbf{Input parameters}                     &       & \textbf{Output parameters}        &                   \\
      Effect size f                                 & 0.25  & Noncentrality parameter $\lambda$  & 12.0             \\
      $\alpha$ error probability                    & 0.05  & Critical F                        & 2.737             \\
      Power ($1-\beta$ error probability)           & 0.8   & Numerator df                      & 3.0               \\
      Number of groups                              & 1     & Denominator df                    & 69                \\
      Number of measurements                        & 4     & \textbf{Total sample size}        & \textbf{24}       \\
      Correlation among representative measures     & 0.5   & Actual power                      & 0.817             \\
      Nonsphericity correction $\epsilon$           & 1     &                                   &                   \\
      \bottomrule
      \end{tabular}%
    \label{tab_GPower}%
  \end{table*}%

%% file: tab_UserLevels.tex
\begin{table*}[h!]
  \centering
  \caption{Definition and distribution of the study participants' user levels based on their professional experience.}
    \begin{tabular}{cclccc}
      \toprule
          &       &       & \multicolumn{3}{c}{\textbf{Professional experience}} \\
\cmidrule{4-6}    \\
\textbf{User level} & \textbf{Number of} & \textbf{Description} & \textbf{Subjective evaluation on} & \textbf{Subjective vehicle} & \textbf{Subjective evaluation} \\
\textbf{ } & \textbf{participants} & \textbf{ } & \textbf{the driving simulator} & \textbf{evaluation} & \textbf{of steering systems} \\
    \midrule
    1     & 8     & Normal driver & \cmark   & \xmark    & \xmark \\
    2     & 14    & Vehicle expert & \cmark   & \cmark   & \xmark \\
    3     & 8    & Steering expert & \cmark   & \cmark   & \cmark \\
    \bottomrule
    \end{tabular}%
  \label{tab_UserLevels}%
\end{table*}%

%% file: tab_MCA.tex
\begin{table*}[h!]
    \centering
    \caption{Motion cueing parameters}
      \begin{tabular}{lccc}
      \toprule
      \multicolumn{1}{p{11.285em}}{\textbf{Degree of Freedom}} & \multicolumn{1}{p{2.5em}}{\textbf{Gain}} & \multicolumn{1}{p{7.5em}}{\textbf{High-pass cut-off frequency (Hz)}} & \multicolumn{1}{p{7.5em}}{\textbf{Low-pass cut-off frequency (Hz)}} \\
      \midrule
      Surge (translation in X)  & 0.5   & 0.5   & 50 \\
      Sway (translation in Y) & 0.5   & 0.25   & 50 \\
      Heave (translation in Z) & 0.5   & 0.5   & 50 \\[0.2cm]
      Roll (rotation around X)  & 0.7   &  -  & 10 \\
      Pitch (rotation around Y) & 0.7   &  -  & 10 \\
      Yaw (rotation around Z) & 0.5   & 0.25   & 50 \\
      \bottomrule
      \end{tabular}%
    \label{tab_MCA}%
  \end{table*}%

%% file: tab_MeanVar_Vehicle.tex
\begin{table*}[h]
    \centering
    \caption{Mean, variance, minimum and maximum values of absolute BI evaluations (Higher means better) split by variant.}
      \begin{tabular}{llccccccc}
      \toprule
      \textbf{ } & \textbf{Variant} & \textbf{SWT on-center} & \textbf{SWT off-center} & \textbf{SWT gradient} & \textbf{Centre feel} & \textbf{Road contact} & \textbf{Low-frequency} & \textbf{High-frequency} \\
        & & & & & & & \textbf{road feedback} & \textbf{road feedback} \\
      \midrule
      \textbf{Mean}         & Reference & 7.89  & 7.89  & 7.41  & 8.11  & 7.74  & 7.63	&	7.8     \\
                            & aVDS V1   & 7.94  & 7.61  & 7.52  & 7.87  & 6.02  & 6.37	& 6.22      \\
                            & aVDS V2   & 7.43  & 7.13  & 6.98  & 7.67  & 5.35  & 5.3	&	4.96    \\
                            & aVDS V3   & 7.46  & 7.57  & 7.33  & 7.85  & 6.56  & 6.43	&	6.44    \\[0.2cm]
        \textbf{Variance}   & Reference & 0.872 & 0.776 & 1.04  & 1.24  & 1.37  & 1.51	&	1.74    \\
                            & aVDS V1   & 0.747 & 0.737 & 1.14  & 1.22  & 2.51  & 3.55	&	2.58    \\
                            & aVDS V2   & 1.38  & 1.53  & 1.2   & 1.08  & 4.63  & 3.43	&	5.29    \\
                            & aVDS V3   & 1.25  & 1.26  & 1.06  & 0.958 & 2.33  & 1.4	&	3.35    \\[0.2cm]
        \textbf{Minimum}    & Reference & 6     & 6     & 5     & 6     & 5     & 6     & 	5       \\
                            & aVDS V1   & 7     & 5.5   & 5     & 6     & 2     & 2	    &	2       \\
                            & aVDS V2   & 5     & 5     & 4     & 6     & 1     & 1	    &	1       \\
                            & aVDS V3   & 5     & 5     & 5     & 6     & 3     & 4	    &	3       \\[0.2cm]
        \textbf{Maximum}    & Reference & 10    & 10    & 9     & 10    & 10    & 10	&	10      \\
                            & aVDS V1   & 10    & 9.5   & 10    & 10    & 10    & 10    & 	10      \\
                            & aVDS V2   & 10    & 9.5   & 9     & 10    & 10    & 8	    &	9       \\
                            & aVDS V3   & 10    & 10    & 9     & 10    & 10    & 8.5   &	10      \\
      \bottomrule
      \end{tabular}%
    \label{tab_MeanVar_Vehicle}%
  \end{table*}%

%% file: tab_MeanVar_UserLevel.tex
\begin{table*}[htbp]
    \centering
    \caption{Mean, variance, minimum and maximum values of absolute BI evaluations (Higher means better) split by User Level.}
      \begin{tabular}{lcccccccc}
      \toprule
      \textbf{ } & \textbf{User} & \textbf{SWT on-center} & \textbf{SWT off-center} & \textbf{SWT gradient} & \textbf{Centre feel} & \textbf{Road contact} & \textbf{Low-frequency} & \textbf{High-frequency} \\
        & \textbf{Level} & & & & & & \textbf{road feedback} & \textbf{road feedback} \\
      \midrule
      \textbf{Mean}         &   0     & 7.91  & 7.75  & 7.48  & 8.18  & 6.68  & 6.43  & 6.66 \\
                            &   1     & 7.46  & 7.27  & 7.22  & 7.89  & 6.21  & 6.32  & 6.08 \\
                            &   2     & 7.81  & 7.8   & 7.3   & 7.59  & 6.5   & 6.59  & 6.5 \\[0.2cm]
        \textbf{Variance}   &   0     & 1.45  & 1.49  & 1.81  & 1.43  & 6.17  & 6.44  & 6.78 \\
                            &   1     & 0.839 & 0.989 & 1.05  & 1.06  & 2.98  & 2.57  & 3.76 \\
                            &   2     & 1.09  & 0.853 & 0.659 & 0.846 & 1.74  & 1.12  & 2.56 \\[0.2cm]
        \textbf{Minimum}    &   0     & 6     & 5.5   & 5     & 6     & 1     & 1     & 1 \\
                            &   1     & 5     & 5     & 4     & 6     & 1     & 2     & 1 \\
                            &   2     & 5     & 5     & 5     & 6     & 4     & 5     & 3 \\[0.2cm]
        \textbf{Maximum}    &   0     & 10    & 10    & 10    & 10    & 10    & 10    & 10 \\
                            &   1     & 10    & 9     & 9     & 10    & 10    & 10    & 10 \\
                            &   2     & 9     & 9.5   & 8.5   & 9     & 9     & 10    & 9.5 \\
      \bottomrule
      \end{tabular}%
    \label{tab_MeanVar_UserLevel}%
  \end{table*}%

%% file: fig_MXeval.tex
\begin{figure*}[h!]
    \centering
    \subfloat[]{\includegraphics[width=7in]{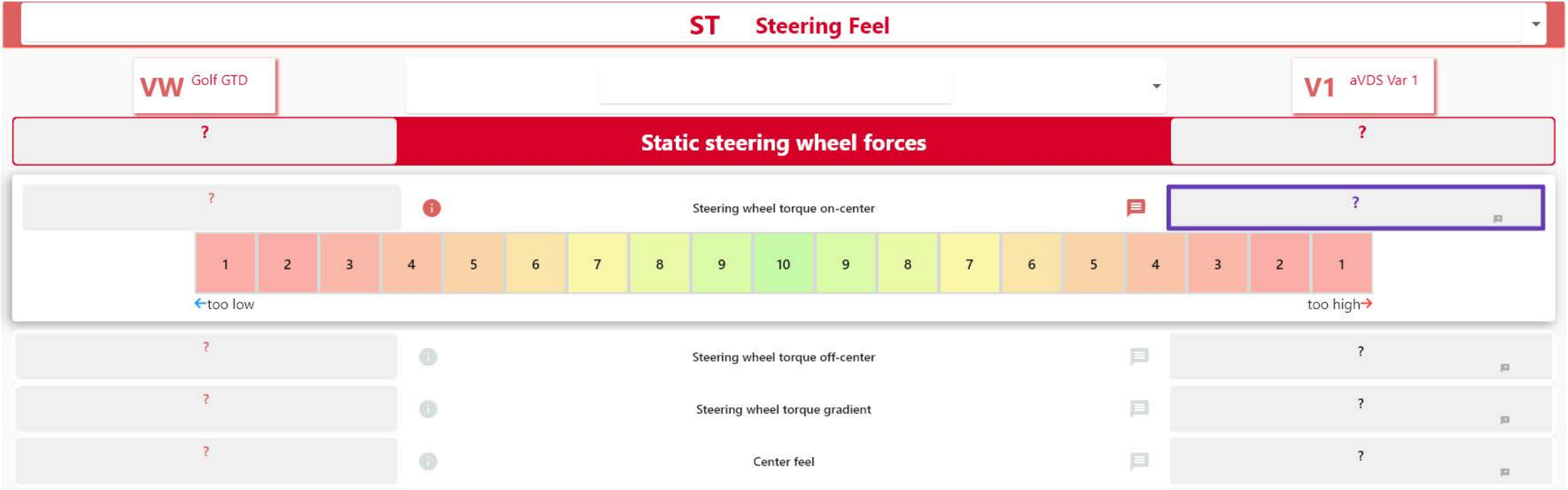}%
    \label{fig_MXeval_BI}}
    \vfil
    \subfloat[]{\includegraphics[width=7in]{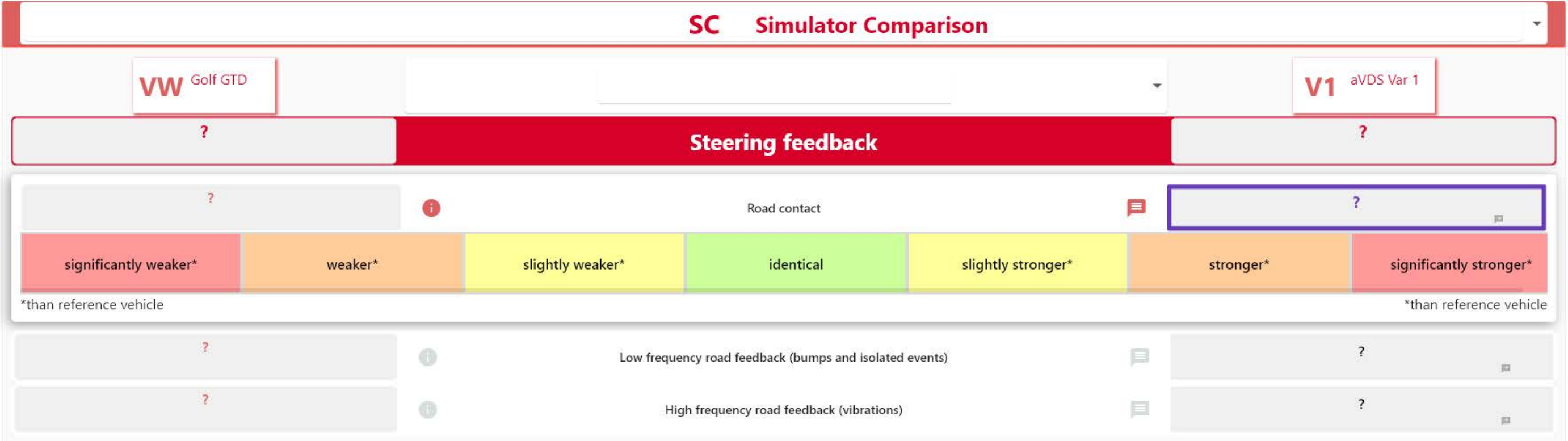}%
    \label{fig_MXeval_Comp}}
    \caption{Screenshot of the questionnaires used for subjective evaluation. \ref{fig_MXeval_BI} BI questionnaire for the evaluation of system performance and \ref{fig_MXeval_Comp} Comparison questionnaire for the evaluation of differences between the simulator variants and the reference vehicle.}
\label{fig_MXeval}
\end{figure*}

%% file: tab_Questionnaire.tex
\begin{table*}[h]
    \centering
    \caption{Questionnaire items in English and German version.}
      \begin{tabular}{llllcc}
        \toprule
      \textbf{Abbreviation} & \textbf{Item}                                 & \textbf{lower}        & \textbf{upper}        & \textbf{linear}               & \textbf{bilinear} \\
      \midrule
      SWT-ON                & Steering wheel torque on-center                               & too low               & too high              & \multirow{2}[0]{*}{\xmark}    & \multirow{2}[0]{*}{\cmark} \\
                            & \textit{Lenkradmomentniveau um die Mitte}                     & \textit{zu niedrig}   & \textit{zu hoch}      &                               &  \\[0.2cm]
      SWT-OFF               & Steering wheel torque off-center                              & too low               & too high              & \multirow{2}[0]{*}{\xmark}    & \multirow{2}[0]{*}{\cmark} \\
                            & \textit{Lenkradmomentniveau off-center}                       & \textit{zu niedrig}   & \textit{zu hoch}      &                               &  \\[0.2cm]
      SWT-G                 & Steering wheel torque gradient on-center                      & too flat              & too steep             & \multirow{2}[0]{*}{\xmark}    & \multirow{2}[0]{*}{\cmark} \\
                            & \textit{Lenkradmomentanstieg aus der Mitte}                   & \textit{zu flach}     & \textit{zu steil}     &                               &  \\[0.2cm]
      CF                    & Centre feel                                                   & too weak              & too strong            & \multirow{2}[0]{*}{\xmark}    & \multirow{2}[0]{*}{\cmark} \\
                            & \textit{Mittengefühl, Zentrierung}                            & \textit{zu schwach}   & \textit{zu stark}     &                               &  \\[0.2cm]
      RC                    & Road contact                                                  & too weak              & adequate              & \multirow{2}[0]{*}{\cmark}    & \multirow{2}[0]{*}{\xmark} \\
                            & \textit{Fahrbahnkontakt}                                      & \textit{zu schwach}   & \textit{ausreichend}  &                               &  \\[0.2cm]
      LF-RF                 & Low-frequency road feedback (bumps and isolated events)       & too weak              & too strong            & \multirow{2}[0]{*}{\xmark}    & \multirow{2}[0]{*}{\cmark} \\
                            & \textit{Fahrbahnrückmeldung niederfrequent (Stöße und Einzelereignisse)} & \textit{zu schwach} & \textit{zu stark} &                          &  \\[0.2cm] 
      HF-RF                 & High-frequency road feedback (vibrations)                     & too weak              & too strong            & \multirow{2}[0]{*}{\xmark}    & \multirow{2}[0]{*}{\cmark} \\
                            & \textit{Fahrbahnrückmeldung hochfrequent (Vibrationen)}       & \textit{zu schwach}   & \textit{zu stark}     &                               &  \\
        \bottomrule
      \end{tabular}%
    \label{tab_Questionnaire}%
  \end{table*}%